\documentclass[aps,prb,twocolumn,showpacs,preprintnumbers,amsmath,amssymb,superscriptaddress]{revtex4-1}

\usepackage{graphicx}
\usepackage{dcolumn}
\usepackage{bm}
\usepackage{color}
\usepackage[dvipsnames]{xcolor}

\newcommand{\be}{\begin{equation}}
\newcommand{\ee}{\end{equation}}
\newcommand{\bea}{\begin{eqnarray}}
\newcommand{\eea}{\end{eqnarray}}

\newcommand{\s}{\sigma}

\newcommand{\up}{\uparrow}
\newcommand{\down}{\downarrow}

\newcommand{\ie}{i.\,e. }

\newcommand{\etal}{{\it et al. }}

\begin{document}

\title{Spin-dependent thermoelectric phenomena in a quantum dot attached to ferromagnetic and superconducting electrodes }

\author{Piotr Trocha}
\email{ptrocha@amu.edu.pl}\affiliation{Faculty of Physics,
Adam Mickiewicz University, 61-614 Pozna\'n, Poland}

\author{J\'ozef Barna\'s}
\affiliation{Faculty of Physics, Adam
Mickiewicz University, 61-614 Pozna\'n, Poland}
\affiliation{Institute of Molecular Physics, Polish Academy of
Sciences, 60-179 Pozna\'n, Poland}

\date{\today}

\begin{abstract}

We investigate transport and thermoelectric properties of hybrid systems based on a single-level quantum dot and one superconducting lead. The other lead is generally normal-metallic ferromagnet. In the latter case single-particle transport is  spin-polarized. Our main interest is in the interplay of Andreev tunneling of Cooper pairs and single-particle tunneling. The latter is responsible for relatively large thermopower and figure of merit due to a diverging density of single-particle states at the superconducting gap edges. System with ferromagnetic and superconducting leads can also reveal spin thermoelectric phenomena. Finite superconducting gap is considered within the BCS theory, and the thermoelectric coefficients are calculated by means
of nonequilibrium Green's function technique within Hartree-Fock like approximation with respect to the intradot Coulomb interaction.

\end{abstract}

\maketitle

\section{Introduction}{\label{Sec:1}}
Thermoelectric properties of nanoscale systems
have been recently attracting much attention, mainly due to expected applications
in nanoelectronic devices for converting  heat to electric energy.
Though such a conversion is known since many decades,
the conversion efficiency needs further improvement. This efficiency is usually measured by
the dimensionless thermoelectric figure of merit  $ZT=\sigma S^2T/\kappa$,
where $T$ stands for temperature, $\sigma$ is the electric
conductivity, $S$ is the Seebeck coefficient (thermopower), whereas
$\kappa$ is the total (phonon plus electronic) thermal conductivity.\cite{ziman,galperinRev,wangRev}
Thus, to enhance the conversion efficiency in bulk materials
one should  increase the thermopower $S$ and electric conductivity $\sigma$
and/or reduce the thermal conductivity $\kappa$. This, however,
turns out to be rather difficult due to the Wiedemann-Franz
law\cite{wied-franz} and Mott relation\cite{mott}.

An alternative way to enhance the thermoelectric efficiency is to use nanoscale systems, where
some enhancement of the figure of merit appears when
reducing dimensions of the system.\cite{hick,beenak,blanter,turek,koch,kubala1,zianni,zhangxm}
This, in turn, may lead to violation of the Wiedemann-Franz
law.\cite{kubala2,murphy} Such systems usually have reduced thermal conductance,
which additionally contributes to the enhancement of $ZT$.\cite{chen}
Following this, interest in thermoelectric phenomena has revived in recent years,
and thermoelectric properties of devices based on quantum dots (QDs) and other
low-dimensional systems have been investigated both
experimentally\cite{reddy,hochbaum,baheti,boukai,schwab} and
theoretically\cite{rego,dubi,swirkowicz,markussen,galperin,lunde,segal,pauly,swirkowiczJP,liu,kuo,szukiewicz}.
It has been shown, for instance,
that electron correlations in the dot strongly influence the thermoelectric
effects in the Kondo
regime\cite{boese,dong,krawiec,sakano,scheibner,franco,yoshida,costi}.
Apart from this,
the thermoelectric phenomena in multiple quantum dots were shown to
reveal a variety of interference effects,
including Fano and Dicke resonances\cite{liuJAP,trocha16}.

When transport is spin polarized due to magnetic electrodes, one may also observe the spin
Seebeck phenomenon.
While the conventional Seebeck effect describes electrical voltage generated by a temperature
gradient, the spin Seebeck effect corresponds to thermally generated spin voltage.
Indeed, such  an effect
was observed experimentally in thin films~\cite{37,uchida} and tunnel junctions,\cite{38,39}
and was also studied theoretically  in transport through other
nanoscale systems.\cite{dubi,swirkowicz,wierzbicki,zberecki}
It is also worth to note that thermoelectric properties of  magnetic multilayered structures remarkably depend on relative alignment
of magnetizations in neighboring magnetic layers\cite{hatami}.

Although, superconductors perfectly conduct electric current and simultaneously are poor thermal conductors, they exhibit very small thermoelectric response~\cite{ginzburg}. In the low-temperature limit, the transport occurs mainly through Andreev states. The problem is that these states are symmetrically located with respect to the Fermi level (due to particle-hole symmetry) and thus no thermopower can be expected.
Therefore, to observe thermoelectricity one has to  break the particle-hole symmetry.
This can be done easily in hybrid systems based on quantum dots, where
breaking of the particle-hole symmetry can be achieved by tuning external gate voltage. Moreover, the particle-hole asymmetry in such systems, and thus the magnitude and sign of the thermoelectric response,  can be electrically tuned  in a controllable manner. Moreover, tuning the dot's energy level $\varepsilon_d$ above the superconducting gap $\Delta$ allows tunneling of hot electrons from a normal metal electrode to empty quasiparticle states in the superconductor (above $\Delta$), which results in a finite thermopower. When the dot's level is located below the gap ($\varepsilon_d < \Delta$), the cold electrons from filled quasiparticle states can tunnel to the empty states of a normal metal, contributing  to the thermopower as well. Thus, by a proper tuning of the gate voltage applied to the dot both magnitude and sign of the thermopower can be manipulated.

Thermoelectric effects in QD-based hybrid structures have been rather weakly investigated. Much more attention has been paid to junctions of normal metals and superconductors, working as refrigerator~\cite{leivo,anghel,giazottoRev,pekola2,pekolaRev}, single-electron radio-frequency cooler~\cite{pekola1}, heat transistor~\cite{saira} and diode~\cite{fornieri2}, or brownian refrigerator~\cite{peltonenPRB14} bearing resemblance to a Maxwell's demon~\cite{koski}. Substantial enhancement of the performance and power of cooling has been achieved in cascade electronic refrigerator~\cite{gomezAPL}. Furthermore, Mazza \etal~\cite{mazza} have shown that electronic heat and charge
currents in a multi-terminal device with
normal and superconducting leads can flow in different reservoirs, which can be considered as heat-charge current
separation.
Other devices like a
heat engine consisting of a voltage-biased
Josephson junction coupled to two resonant cavities~\cite{hofer}, or thermal modulator providing a full control of the phase engineering of electronic heat current at the nanoscale~\cite{fornieri}
have been also suggested.
Apart from this, Kleeorin \etal have demonstrated that Josephson junction with a quantum dot placed between two superconductors can reveal highly
tunable thermoelectric properties and a significant thermal response~\cite{Kleeorin}.

Although vast literature on transport properties of QDs coupled to normal and superconducting leads already exist,\cite{yeyati,deacon,fazioPRL98,sunQDSC,WangPRB,Zeng,donabidowicz07,domanski08,panSC2,zhangSC,tanakaSC,lopezSC,pengPRB05,governale08,wojcik} only a few papers address thermoelectric properties~\cite{krawiecSC,hwang1,hwang2}. Moreover, mainly, the Andreev transport regime was  investigated in three-terminal devices, where thermoelectricity appeared between normal metal leads in the presence  of an additional  superconducting contact~\cite{wysokinskiJPCM,wojcik,caoSC,michalek}. Remarkably less attention has been paid to transport though QDs coupled to ferromagnetic and superconducting leads.\cite{feng03,cao,pengZhang,wysokinskiJPCM,csonka}
Thermoelectric properties of such systems are very weakly explored,~\cite{hwang1,hwang2} though recent observation of
spin-dependent thermoelectric currents in superconductor-ferromagnet tunnel junctions in a high magnetic field has shown the Seebeck
coefficient much larger than commonly found in metallic structures~\cite{kolenda}.

Therefore, in this paper we consider thermoelectric properties of a single-level quantum dot attached to one ferromagnetic metallic and one superconducting lead.
We include Coulomb correlations in the dot~\cite{krawiecSC,hwangNJP}, which is often dropped for the sake of simplicity~\cite{hwang1,hwang2,valentini}.
The basic thermoelectric characteristics are derived using
 the nonequilibrium Green function method. Since the Coulomb correlations are
taken in the Hartree-Fock approximation, the formalism is
relevant for temperatures above the Kondo temperature.
We consider a general situation, where charge transport
appears due to Andreev tunneling of Cooper pairs as well as due to
over-barrier tunneling of single electrons. The former transport
dominates at low temperatures, while the latter becomes relevant
when the thermal energy is comparable to
the superconducting gap. Since Andreev tunneling itself does not give rise to thermal conductance and thermopower in the linear response regime, the latter tunneling process become then crucial.
However, we show that Andreev tunneling has a remarkable impact on thermoelectric properties when the single-electron tunneling processes are present.
When single-particle transport is spin-polarized (Andreev current is not spin polarized), the system may also reveal
spin thermoelectricity. Moreover, since the density of states at the edges of the superconducting gap is very high and rapidly changes with energy, the thermopower as well as the thermoelectric efficiency, both conventional and spin, can achieve relatively high values.

In Section II we describe the model of the system and present a general background on the
thermoelectric properties. In Section III we
present the corresponding numerical results for the QDs with large on-site Hubbard parameter, so these are only singly-occupied. In section IV we consider the role of Coulomb interaction in the dot. Finally,
 Section V includes summary and general conclusions.

\section{Theoretical description}

\subsection{Model}

\begin{figure}
\includegraphics[width=0.8\columnwidth,angle=0]{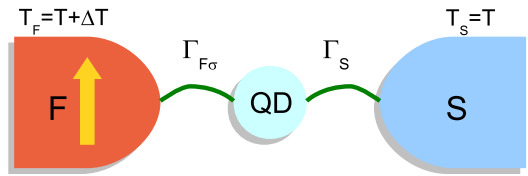}
\caption{\label{Fig:1}
\label{Fig:1} Schematic picture of quantum-dot based thermoelectric device.
Quantum dot is coupled to the left metallic (ferromagnetic) lead with the strength
$\Gamma_{F\sigma}$ and to right superconducting electrode with the strength
$\Gamma_S$. In the case of ferromagnetic lead the two spin channels are coupled to
the QD with different strengths. We assume the left lead is kept at higher temperature than
the right one which is intuitively illustrated by colors.
}
\end{figure}

The system under consideration consists of a single-level quantum dot attached
to one metallic ferromagnetic and one superconducting lead, as shown schematically in Fig.1.
Such a system can be modeled by the following general Hamiltonian:
\begin{equation}\label{Eq:1}
H=H_{c}+H_{QD}+H_{T}.
\end{equation}
The first term, $H_{c}$, represents Hamiltonian of the external leads: $H_{c}=H_L +H_R$. Here, $H_L=\sum_{{\mathbf
k}}\sum_\sigma\varepsilon_{{\mathbf k}L\sigma}
     c^\dag_{{\mathbf k}L\sigma}c_{{\mathbf k}L\sigma}$
describes the ferromagnetic lead in the non-interacting
quasi-particle approximation, assumed to be on the left (L) side, with $\varepsilon_{{\mathbf k}L\sigma}$ representing the spin-dependent electronic spectrum.  In turn, $H_R$ stands for the BCS
Hamiltonian of the superconducting (right) lead in the mean field approximation,
\begin{eqnarray}\label{Eq:2}
H_{R}&=&\sum_{{\mathbf
k}}\limits\sum_\sigma\limits\varepsilon_{{\mathbf k}R}
     c^\dag_{{\mathbf k}R\sigma}c_{{\mathbf k}R\sigma}
     \nonumber\\
     &+&
     \sum_{{\mathbf k}}\limits\left(\Delta^{\ast} c_{{\mathbf
k}R\downarrow}c_{{-\mathbf k}R\uparrow}+\Delta c^{\dag}_{{-\mathbf
k}R\uparrow}c^{\dag}_{{\mathbf k}R\downarrow}\right),
\end{eqnarray}
where $\varepsilon_{{\mathbf k}R}$ denotes the relevant
single-particle energy spectrum (independent of spin orientation) and  $\Delta=|\Delta|e^{i\Phi}$ stands
for the  order parameter of the superconductor. Accordingly, $|\Delta|$
denotes the superconducting gap, whereas $\Phi$ is the relevant
phase. In the problem considered here the phase factor $\Phi$ is
irrelevant as there is only one superconducting lead and therefore it will be omitted in the following.

The second term in Eq.(1) describes a single-level quantum dot and has the form,
\begin{equation}\label{Eq:2}
\hat{H}_{QD}=\sum_{\sigma}\varepsilon_{\sigma}d^\dag_{\sigma}d_{\sigma}+
   Un_{\uparrow}n_{\downarrow},
   \end{equation}
where $\varepsilon_{\sigma}$ denotes the dot's energy level, which can be spin-dependent ($\sigma=\up, \down$) in a general case. The parameter $U$
stands for the Coulomb energy of two electrons residing in the dot.

The last term of the Hamiltonian (1) describes tunneling of electrons
between the leads ($L,R$) and the dot,
\begin{eqnarray}\label{Eq:4}
H_T=\sum_{\mathbf{k}\sigma}\limits\sum_{i=L,R}
   \limits (V_{\mathbf{k}\sigma}^i c^\dag_{\mathbf{k}i\sigma}d_{\sigma}+\rm
   H.c.),
\end{eqnarray}
with $V_{\mathbf{k}\sigma}^i$, for $i=L,R$,
denoting the relevant tunneling matrix elements between the dot
and $i$-th lead.
Furthermore, we assume that these matrix elements
are independent of $\mathbf{k}$ and $\sigma$,
$V_{\mathbf{k}\sigma}^i \equiv V^i$.
Coupling of the dot to the ferromagnetic, $i=L$, lead
can be then parameterized by $\Gamma_{L}^{\sigma}=2\pi|V^{L}|^2\rho_{L}^\sigma$,
where $\rho_{L}^\sigma$ is the spin-dependent density of states in the ferromagnetic lead.
Within the wide band approximation these coupling parameters become
independent of energy and can be considered as constant.
Introducing the spin polarization of the ferromagnetic lead,
$p=(\rho_{L}^+-\rho_{L}^-)/(\rho_{L}^++\rho_{L}^-)$,
where $\rho_{L}^+$ ($\rho_{L}^-$) is the spin majority (minority) density of states,
the coupling parameters can be written in the form,
$\Gamma_{L}^{\sigma} = (1+ \tilde{\sigma} p) \Gamma_L$,
with $\Gamma_L = (\Gamma_L^\up + \Gamma_L^\down)/2$ and $\tilde{\sigma} =1 (-1)$ for
$\sigma =\uparrow (\downarrow )$. We assume $\uparrow$ corresponds to spin orientation of
spin-majority electrons.  Similarly, coupling to the superconducting electrode in the corresponding normal
state is denoted as $\Gamma_{R}=2\pi|V^{R}|^2\rho_{R}$, and is independent of spin orientation and of energy. This coupling becomes modified in the
superconductiong state, as will be described in the following.

\subsection{Charge current}

Charge current $J$ flowing  in a biased system from left to right can be calculated from the following formula~\cite{Zeng}:
\begin{eqnarray}\label{Eq:3}
  J &=&\sum_\sigma J_\sigma \equiv\frac{ie}{2\hbar}\sum_\sigma\int\frac{d\varepsilon}{2\pi}
  \left\{[\mathbf{\Gamma}^{\sigma}_{L}-\mathbf{\Gamma}^{\sigma}_{R}]
  \mathbf{G}^{<}_{\sigma}(\varepsilon)\right.
  \nonumber\\
  &+&
  \left.[\mathbf{f}_L^\sigma(\varepsilon)\mathbf{\Gamma}^{\sigma}_{L}-
   \mathbf{f}_R(\varepsilon)\mathbf{\Gamma}^{\sigma}_{R}]
   [\mathbf{G}^{r}_{\sigma}(\varepsilon)-\mathbf{G}^{a}_{\sigma}(\varepsilon)]\right\}_{11},
\end{eqnarray}
where $\mathbf{G}_\sigma^{r,a,<}$ are the retarded, advanced, and lesser Green's functions in the matrix form,
while $\mathbf{\Gamma}^{\sigma}_{\beta}$ (for $\beta =L,R$)
denote the couplings to the leads (also in the matrix form). More specifically,
\begin{equation}\label{Eq:x}
  \mathbf{\Gamma}^{\sigma}_{L}=
    \left(%
\begin{array}{cccc}
  \Gamma_{L}^{\sigma} & 0 \\
  0 & \Gamma_{L}^{\bar{\sigma}}\\
\end{array}%
\right)
\end{equation}
for the coupling to the magnetic lead, where $\bar{\sigma}\equiv -\sigma$, and
\begin{equation}\label{Eq:y}
  \mathbf{\Gamma}^{\sigma}_{R}=
\rho_R(\varepsilon)\Gamma_R
  \left(%
\begin{array}{cccc}
  1 & -\tilde{\sigma}\frac{\Delta}{|\varepsilon|}  \\
  -\tilde{\sigma}\frac{\Delta}{|\varepsilon|} & 1 \\
\end{array}%
\right),
\end{equation}
for the coupling to the right lead in the superconducting state, with
\begin{equation}\label{Eq:z}
\rho_R(\varepsilon)=\frac{|\varepsilon|\theta(|\varepsilon|-\Delta)}
{\sqrt{\varepsilon^2-\Delta^2}}
\end{equation}
denoting the superconducting density of states.
Finally, $\mathbf{f}_{L}^\sigma (\varepsilon)$ describes the
Fermi-Dirac distribution in the ferromagnetic lead, assumed to be spin dependent,
\begin{equation}\label{Eq:14}
\mathbf{f}_{L}^\sigma =
\left(%
\begin{array}{cccc}
  f(\varepsilon-eV_\sigma) & 0 \\
  0 & f(\varepsilon+eV_{\bar{\sigma}}). \\
\end{array}%
\right),
\end{equation}
while $\mathbf{f}_{R}=f(\varepsilon)\,{\rm diag}(1,1)$ is the Fermi distribution for the
superconducting lead assumed to be independent of the spin orientation.

To derive the lesser Green's function we apply the Keldysh relation,
\begin{eqnarray}
\mathbf{G}^{<}_{\s}=\mathbf{G}^{r}_{\s}\mathbf{\Sigma}^{<}_{\s}\mathbf{G}^{a}_{\s},\label{Eq:4}
\end{eqnarray}
whereas the lesser self-energy can be obtained from the following
formula:
\begin{eqnarray}
\mathbf{\Sigma}^{<}_{\s}=\mathbf{\Sigma}^{<}_{L\s}+\mathbf{\Sigma}^{<}_{R\s}=
i(\mathbf{f}_L^\sigma(\varepsilon)\mathbf{\Gamma}_{L}^\s+
   \mathbf{f}_R(\varepsilon)\mathbf{\Gamma}_{R}^\s),\label{Eq:5}
\end{eqnarray}
which is valid for interactions
taken in the mean field approximation.

The retarded Green's function $\mathbf{G}^{r}_{\s}(\varepsilon)$ has
been obtained from the Dyson equation,
\begin{equation}\label{Eq:9}
\mathbf{G}^{r}_{\s}=[(\mathbf{g}^r)^{-1}_{\s}+\mathbf{\Sigma}^r_{\s}]^{-1},
\end{equation}
with $\mathbf{g}^r_{\s}$ denoting the retarded
Green's function of the dot isolated from the leads, and
$\mathbf{\Sigma}^r_\sigma =\mathbf{\Sigma}^r_{L\sigma}+\mathbf{\Sigma}^r_{R\sigma}$ representing the retarded self-energy which describes
interaction between the QD and electrodes. The retarded Green's
function  $\mathbf{g}^{r}_{\s}$ has been derived from the relevant
equation of motion. In turn,
the retarded
self-energy due to coupling to the ferromagnetic lead, $\mathbf{\Sigma}^r_{L\s}$,
taken in the wide band approximation acquires the form,
\begin{equation}\label{Eq:11}
  \mathbf{\Sigma}^{r}_{L\s}=
  -\frac{i}{2}
  \left(%
\begin{array}{cccc}
  \Gamma_{L}^{\sigma} & 0 \\
  0 & \Gamma_{L}^{\bar{\sigma}}\\
\end{array}%
\right),
\end{equation}
whereas the self-energy due to coupling to the right
(superconducting) lead takes the form
\begin{equation}\label{Eq:12}
  \mathbf{\Sigma}^{r}_{R\sigma}=
  -\frac{i}{2}
  \tilde{\rho}_R(\varepsilon)
  \left(%
\begin{array}{cccc}
  \Gamma_{R} & -\tilde{\sigma}\frac{\Delta}{|\varepsilon|}\Gamma_{R}  \\
  -\tilde{\sigma}\frac{\Delta}{|\varepsilon|}\Gamma_{R} & \Gamma_{R} \\
\end{array}%
\right),
\end{equation}
where $\tilde{\rho}_R(\varepsilon)$ denotes the dimensionless  modified
BCS density of states in the superconductor,
\begin{equation}\label{Eq:13}
\tilde{\rho}_R(\varepsilon)=\frac{|\varepsilon|\theta(|\varepsilon|-\Delta)}
{\sqrt{\varepsilon^2-\Delta^2}} +
\frac{\varepsilon\theta(\Delta-|\varepsilon|)}
{i\sqrt{\Delta^2-\varepsilon^2}}.
\end{equation}
It is worth noting that the coupling matrix to the
right lead, $\mathbf{\Gamma}_{R}^\sigma$, is proportional to the
nonmodified BCS density of states, see Eq.(7), while the self-energy (14) is proportional to
the modified BCS density of states
$\tilde{\rho}_R(\varepsilon)$.

Making use of the above equations and taking into account the
identity,
$\mathbf{G}^{r}_{\s}-\mathbf{G}^{a}_{\s}=-i\mathbf{G}^{r}_{\s}\mathbf{\Gamma}^\s\mathbf{G}^{a}_{\s}$,
with $\mathbf{\Gamma}^\s=\mathbf{\Gamma}_{L}^\s+\mathbf{\Gamma}_{R}^\s$,
the current expression (\ref{Eq:3}) simplifies to the
Landauer-like formula. When the bias voltage is applied to the
system, with the electrochemical potential of the ferromagnetic
lead equal to $\mu_L^\sigma =eV_\sigma$ and electrochemical potential of the
superconducting electrode equal to zero, $\mu_R=0$, the current can be written as a sum of
current $J^S$ due to single particles and current $J^A$ due to Andreev transmission,
\begin{equation}\label{Eq:6}
  J=\sum_{\sigma} J_\sigma = \sum_{\sigma} (J^S_\sigma+J^{A}_\sigma),
\end{equation}
with
\begin{equation}
J^S_\sigma =\frac{e}{h} \int d\varepsilon [f_L(\varepsilon-eV_\sigma)-f_R(\varepsilon)]T^S_\s(\varepsilon),
\label{Eq:7}
\end{equation}
\begin{equation}
J^A_\sigma =\frac{e}{h} \int d\varepsilon [f_L(\varepsilon-eV_\sigma)-f_L(\varepsilon+eV_{\bar{\sigma}})]T^A_\s(\varepsilon)
.\label{Eq:8}
\end{equation}
with the corresponding transmission functions $T^S_\s(\varepsilon)=[\mathbf{G}^{r}_\sigma\mathbf{\Gamma}_{R}^\sigma\mathbf{G}^{a}_\sigma\mathbf{\Gamma}_{L}^\sigma]_{11}$ and $T^A_\s(\varepsilon)=G^r_{12\s}[\mathbf{\Gamma}_{L}^\sigma\mathbf{G}^{a}_{\s}\mathbf{\Gamma}_{L}^{\s}]_{21}$.

\subsection{Thermoelectric coefficients}

Now we consider thermoelectric coefficients, and will distinguish between the situations with and without spin voltage.
Consider first the latter case, i.e. the absence of  spin voltage. This corresponds to the conventional thermoelectricity.

\subsubsection{Conventional thermoelectricity}

Assume now independent of spin chemical potential in the ferromagnetic lead, $\mu_L^\sigma =\mu_L$, i.e. $V_\sigma=V$.
Charge current is then given by Eqs (16-18) where we put $V_\sigma=V$. In turn, the heat current $J_Q $ flowing out of the ferromagnetic lead can be written as
\be
J_Q = \sum_\sigma (J_{Q\sigma}^S +J_{Q\sigma}^A),
\ee
with $J_{Q\sigma}^S$ denoting the contribution due to single particle tunneling in the spin channel $\sigma$,
\begin{equation}
J^S_{Q\sigma}=\frac{1}{h} \int d\varepsilon (\varepsilon-\mu_L)[f_L(\varepsilon-eV)-f_R(\varepsilon)]T^S_\s(\varepsilon),
\label{Eq:7}
\end{equation}
and $J_{Q\sigma}^A$ denting the contribution due to Andreev tunneling,
\begin{equation}
J_{Q\sigma}^A=-V\frac{e}{h}\int d\varepsilon [f_L(\varepsilon-eV)-f_L(\varepsilon+eV)]T^A_\s(\varepsilon).
\end{equation}

In the linear response regime the equations for charge current (16 to 18) and heat current (19 to 21)) can be written as
\begin{equation}\label{Eq:4}
J=\sum_\sigma J_{\sigma} \equiv e \sum_\sigma L_{11}^\s
 \delta \mu+\frac{e}{T}\sum_\sigma L_{12}^\s\delta  T,
\end{equation}
\begin{equation}\label{Eq:6}
J_Q=\sum_\sigma J^Q_\sigma \equiv \sum_\sigma L_{21}^\s\delta
\mu +\frac{1}{T} \sum_\sigma L_{22}^\s\delta  T,
\end{equation}
where $\delta  T$ is the difference in temperatures of the leads,
and $\delta \mu=e\delta  V$, with $\delta  V$ being the voltage
drop between the two electrodes. According to our
assumption, the chemical potential and temperature of the left
electrode is $\delta  \mu$ and $T +\delta  T$, respectively,
whereas of the right electrode are $\mu =0$ and $T$. Note, that in
the linear response regime both $\delta \mu$ and $\delta T$ are infinitesimally small. Thus,
$J_Q$ is then the heat current flowing from the left to right lead.
In Eqs.~(\ref{Eq:4}) to (\ref{Eq:6}) the integrals $L_{nm}^\s$
($n,m=1,2;\sigma =\uparrow, \downarrow$) are of the
form
\begin{subequations}
\begin{align}
L_{11}^\s=-\frac{1}{h}\int  d\varepsilon
\frac{\partial
f}{\partial\varepsilon}\left[ 2T_{\sigma}^A(\varepsilon)+T_{\sigma}^S(\varepsilon)\right],
\label{Eq:8a} \\
L_{12}^\s=L_{21}^\s=-\frac{1}{h}\int  d\varepsilon
(\varepsilon-\mu)\frac{\partial
f}{\partial\varepsilon}T_{\sigma}^S(\varepsilon),
\label{Eq:8b}\\
L_{22}^\s=-\frac{1}{h}\int  d\varepsilon
(\varepsilon-\mu)^2\frac{\partial
f}{\partial\varepsilon}T_{\sigma}^S(\varepsilon).
\end{align}
\end{subequations}
Note, Andreev tunneling does not contribute to heat current in the linear response regime (and thus does not contribute to the heat conductance).

The thermopower $S$ is defined as the the ratio of the voltage
drop $\delta  V$ generated by the temperature difference $\delta
T$, $S=-\delta  V/\delta  T$, taken in the absence of charge
current, $J=0$. Thus, taking into account  Eq.(\ref{Eq:4}) and
Eq.(\ref{Eq:6}), one obtains the following formula for the Seebeck
coefficient:
\begin{equation}\label{Eq:8}
S = -\left[\frac{\delta  V}{\delta
T}\right]_{J=0}=\frac{1}{eT}\frac{\sum_\sigma
L_{12}^\s}{\sum_\sigma L_{11}^\s}.
\end{equation}
Similarly, the charge  conductance
$G$
can be expressed in terms of the integral (24)a  as~\cite{ziman}
\begin{equation}\label{Eq:9}
G=e^2\sum_\sigma L_{11}^\s,
\end{equation}
while the thermal conductance can be written as
\begin{equation}\label{Eq:11}
\kappa=\frac{1}{T}\left(\sum_\sigma L_{22}^\s-\frac{[\sum_{\s}
L_{12}^{\s}][\sum_{\s}L_{21}^\s]} {\sum_\sigma L_{11}^\s}\right).
\end{equation}
Note, the thermal conductance is defined by the heat current in the absence of charge current. Finally,
the figure of merit $ZT$ is given as  $ZT=GS^2T/\kappa$.

\subsubsection{Spin thermoelectricity}

Here, we consider the situation, when spin voltage due to spin accumulation in
the external ferromagnetic lead becomes relevant, eg. due to long spin
relaxation time or when an external spin dependent bias is applied
to the system. Then, we have to take into account spin
splitting of the chemical potential in the leads. In a general
case, temperature may also be different in different spin
channels. However, we neglect this assuming $T$ independent of
$\sigma$. This may be justified as the energy relaxation time is
much shorter than the spin relaxation one.
Now spin thermoelectricty becomes relevant and to determine the corresponding thermoelectric efficiency
we will also need the spin conductance.
The, charge, spin and heat currents can
be expressed by formulas similar to Eqs (\ref{Eq:4}) to
(\ref{Eq:6}), but with $\delta\mu$ being explicitly spin dependent,
$\delta\mu\to \delta\mu_\sigma$, and $L_{nm}^\s$
including now derivative of spin dependent Fermi distribution
function.

Since the bias is now spin dependent, the difference in chemical
potentials, $\delta\mu_\sigma$, in the spin channel $\sigma$ can
be written as
\begin{equation}\label{Eq:13}
\delta\mu_\sigma = e\delta V_\sigma =e(\delta V+\hat{\sigma}
\delta V^s),
\end{equation}
where $\delta V$ is the conventional voltage and $\delta V^s$ is the spin
voltage~\cite{swirkowicz}. Of course, $\delta V^s=0$ in the absence of
spin accumulation.

Charge and spin currents can be written as $J=G\delta V +(2e/\hbar
)G^s\delta V^s$, where $G$ and $G^s$ are the linear charge and
spin conductances. In turn,
the spin current can be written as $J^s=G^s\delta
V+(\hbar/2e)G\delta V^s$.

The thermopower can be calculated on the condition of
vanishing simultaneously both spin current and charge current, or
equivalently on the condition of vanishing charge current in each
spin channel. As a result, one can define spin dependent
thermopower as
\begin{equation}\label{Eq:16}
S_\sigma =-\frac{\delta V_\sigma}{\delta T}= \frac{L_{12}^\s}{
eTL_{11}^\s}.
\end{equation}
Equivalently,  one can define spin thermopower $S^s$
\begin{equation}\label{Eq:17}
S^s =-\frac{\delta V^s}{\delta T}= \frac{1}{2}(S_\uparrow
-S_\downarrow)=\frac{1}{2eT}\left(\frac{L_{12}^\up}{
L_{11}^\down}-\frac{L_{12}^\up}{ L_{11}^\down}\right)
\end{equation}
in addition to the usual thermopower
\begin{equation}\label{Eq:18}
S =-\frac{\delta V}{\delta T}= \frac{1}{2}(S_\uparrow
+S_\downarrow)=\frac{1}{2eT}\left(\frac{L_{12}^\up}{
L_{11}^\up}+\frac{L_{12}^\down}{ L_{11}^\down}\right).
\end{equation}
In turn, the heat conductance is then given by
\begin{equation}\label{Eq:19}
\kappa=\sum_\sigma \kappa_\sigma \equiv
\frac{1}{T}\sum_\sigma\left(L_{22}^\s-\frac{L_{12}^\s L_{21}^\s}{L_{11}^\s}\right),
\end{equation}
where now both charge and spin currents are assumed to vanish. Finally, the spin figure of merit
can be defined as $Z_sT=(2|e|/\hbar )(|G_s|S_s^2T/\kappa )$.

Below we will present some numerical results obtained from the above formulas.  All energy quantities in the following will be expressed in the units of $\Delta$. For the superconducting
lead, the couplings are: $\Gamma_{R}^\s=\Gamma_R=r\Gamma$, where the parameter $r$
takes into account asymmetry in the couplings of the dot to left and right leads, so
$\Gamma_L=\Gamma$. In numerical calculations we assume  $\Gamma/\Delta=0.1$ and spin degenerate dot's energy level,
$\varepsilon_\up=\varepsilon_\down\equiv\varepsilon$, if not stated otherwise. We will consider separately the case of zero Coulomb energy on the dot, and finite value of the  parameter $U$. Let us begin with the former case.

\section{Numerical results for $U=0$}

In this section we present numerical results obtained for $U=0$. To understand better the results in a general case (magnetic leads) it is instructive and also desirable to consider first the limit of nonmagnetic leads. Therefore, we first consider briefly the case of $p=0$.

\subsection{Case of nonmagnetic lead, $p=0$}

\begin{figure}
\includegraphics[width=\columnwidth,angle=0]{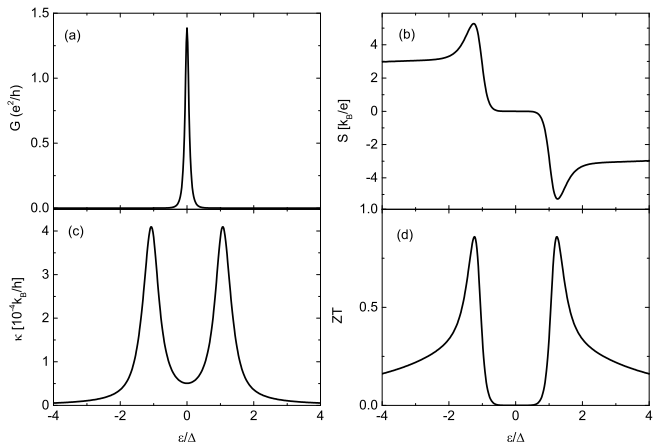}
\caption{\label{Fig:2}
Transport and thermoelectric coefficients:
(a) electrical conductance, (b) thermopower, (c) thermal conductance, and (d) figure of merit,
calculated as a function of the dot's level energy for $k_BT/\Delta
=0.1$, $\Gamma/\Delta =0.1$, $U=0$, $p=0$, and
$r=1$.}
\end{figure}

In Fig.~\ref{Fig:2} we show the linear electric conductance $G$, thermopower $S$, electronic contribution to the heat conductance $\kappa$, and figure of merit $ZT$,
calculated as a function of the dot's level energy $\varepsilon /\Delta$ for $k_BT/\Delta
=0.1$, $\Gamma/\Delta =0.1$, $U=0$, $p=0$, and
for a symmetric coupling to the left and right leads,  $r=1$.

The electrical conductance $G$ displays a narrow peak centered at the Fermi level, $\varepsilon =0$.
Generally, the conductance includes a contribution from the Andreev transmission of Cooper pairs and another contribution from single-particle tunneling to states above (below) the superconducting gap.
Since we consider the linear response limit, and $k_BT/\Delta << 1$, the conductance $G$ in Fig.~\ref{Fig:2}a originates mainly from the
Andreev tunneling. The contribution from single particle tunneling appears at larger values of $\varepsilon /\Delta \sim k_BT /\Delta \sim 1$, as shown below. Since $k_BT/\Delta \ll 1$, this contribution is much smaller than that due to Andreev tunneling and is not well visible in Fig.~\ref{Fig:2}a.

The thermal conductance $\kappa$ is shown in Fig.~\ref{Fig:2}c as a function of the dot's level energy. To understand this dependence it is worth to remind that the thermal conductance is determined at the condition of zero charge current, and to note that Andreev tunneling does not contribute to heat current in a thermally biased system. The thermal conductance  reveals two peaks associated with single electron tunneling to states above (below) the gap in the superconducting lead.
As one might expect, these peaks appear at $\varepsilon /\Delta \sim \pm 1$. Interestingly, the corresponding peaks in electrical conductance are
hardly seen in Fig.~\ref{Fig:2}a, though they are well seen in the heat conductance. There are several reasons of this. First, contribution to heat current of a single particle tunneling process grows with energy of tunneling electron and is relatively large -- it is of the order of $\Delta$, while contribution to charge current is independent of electron energy. Moreover, the contribution of single electron tunneling processes is obscured in Fig.~\ref{Fig:2}a by a much larger contribution due to Andreev tunneling, while Andreev tunneling does not contribute to heat transfer to the superconductor in the thermally biased system. Thus, though the heat conductance due to single electron tunneling is rather small, it is seen in Fig.~\ref{Fig:2}c.  Apart from this, the temperature assumed in Fig.~\ref{Fig:2} is relatively low, which suppresses the single electron tunneling. To increase relative contribution of single electron tunneling one needs to increase temperature, as will be seen in the following.

The Seebeck coefficient (thermopower) is shown in Fig.~\ref{Fig:2}b. The thermopower vanishes exactly in the particle-hole symmetry point,  $\varepsilon =0$, where the temperature difference does not lead to any charge current. It also vanishes for $\varepsilon$ in the superconducting gap, since the single electron tunneling processes are then suppressed, whereas the temperature gradient does not give rise to Andreev current, as already mentioned above. The thermopower rapidly increases at the edges of the superconducting gap, $|\varepsilon|\sim\Delta$, where the single electron tunneling processes are admitted and contribute to charge current. Since the thermopower is measured at zero charge current, one needs to apply a certain voltage to compensate this charge  current. The compensating current includes also a current due to Andreev tunneling. Maxima in $|S|$ develop in the vicinity of the maxima in the density of single particle states near the superconducting gap edges.
When absolute value of $\varepsilon$ increases further, the thermopower in Fig.~\ref{Fig:2}b decreases very slowly. This behavior appears as a result of a small electrical conductance for large values of $|\varepsilon |$, so the external voltage applied to suppress the current must be relatively large.

The above behavior of the thermopower $S$ and of the electrical $G$ and thermal $\kappa$ conductances leads to the figure of merit $ZT$ shown in Fig.~\ref{Fig:2}d. Note, the phonon contribution to the heat conductance is neglected here.
The figure of merit $ZT$ is equal to zero for $|\varepsilon |$ in the superconducting gap, as the thermopower $S$ is negligible there (see Fig.~\ref{Fig:2}b. Then $ZT$  increases with increasing $|\varepsilon |$, and reaches maxima around the superconducting gap edges. When $|\varepsilon |$ increases further, figure of merit $ZT$ decreases due to suppressed electrical conductance, but this decrease is relatively slow due to the reduced thermal conductance. Interestingly, the maximum value of $ZT$ is of the order of 1.

\begin{figure}[]
\includegraphics[width=\columnwidth,angle=0]{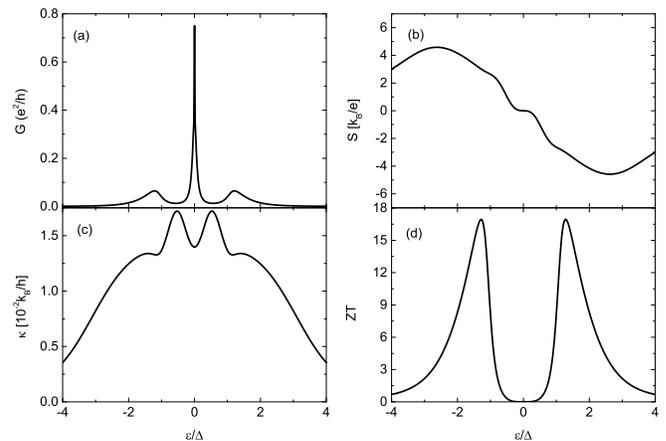}
\caption{\label{Fig:4}
Transport and thermoelectric coefficients:
(a) electrical conductance, (b) thermopower, (c) thermal conductance, and (d) figure of merit,
calculated as a function of the dot's level energy for $k_BT/\Delta
=0.4$, $\Gamma/\Delta =0.1$, $U=0$, $p=0$, and
$r=1$.}
\end{figure}

The results for remarkably higher temperature, $k_BT/\Delta =0.4$, are shown in Fig.~\ref{Fig:4}. Now, the contribution of single-electron
tunneling is much higher, as clearly seen in Fig.~\ref{Fig:4}a, where the side peaks are relatively high, while the central peak due to Andreev tunneling
is reduced. The thermal conductance (Fig.~\ref{Fig:4}c) is enhanced by roughly two orders of magnitude in comparison to that in Fig.~\ref{Fig:2}c,  while the thermopower (Fig.~\ref{Fig:4}b) is
qualitatively and quantitatively similar to the thermopower  in Fig.~\ref{Fig:2}b. In turn, the figure of merit $ZT$ (Fig.~\ref{Fig:4}d) is significantly enhanced, though its dependence on $\varepsilon$ is qualitatively similar to that in Fig.~\ref{Fig:2}d.

\subsection{Case of magnetic leads, $p\neq 0$}

\begin{figure}[]
\includegraphics[width=\columnwidth,angle=0]{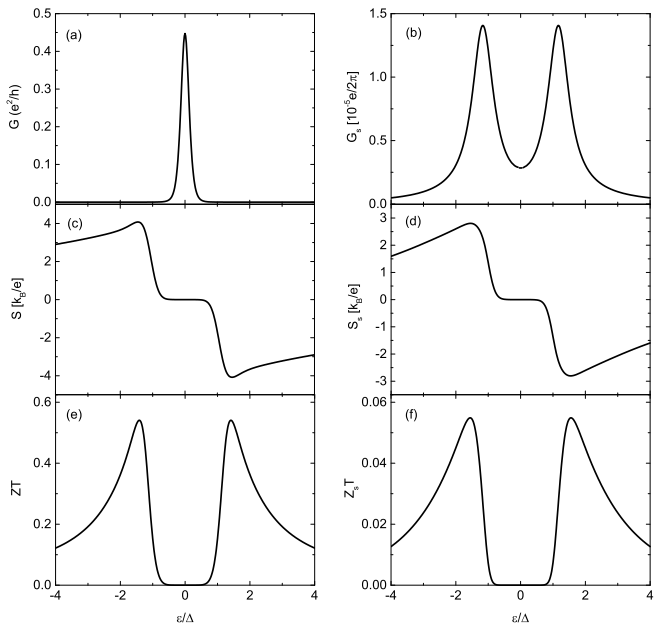}
\caption{\label{Fig:6}
Spin thermoelectric coefficients:
(a) charge conductance, (b) spin conductance, (c) thermopower, (d) spin thermopower,
(e) figure of merit $ZT$, and (f) spin figure of merit $Z_sT$,
calculated as a function of the dot's level energy for $k_BT/\Delta
=0.1$, $\Gamma/\Delta =0.1$, $U=0$, $p=0.9$ and
$r=2$.}
\end{figure}

Let us consider now the situation of spin polarized transport, i.e. the case
when the external metallic lead is ferromagnetic, $p\ne 0$. One can observe then not only the conventional thermopower, but also the spin
thermopower, as introduced in Sec. 2.
The charge and spin conductances, the conventional and spin thermopowers,  as well as the corresponding conventional and spin figures of merit are shown in Fig.~\ref{Fig:6} for $k_BT/\Delta =0.1$, $r=2$  and $p=0.9$. As one might expect, the general variation of the charge  conductance with position of the dot's level $\varepsilon$, see Fig.4a,  is qualitatively similar to that in the case of nonmagnetic metallic lead. However, the Andreev contribution is reduced in comparison with the corresponding one for $p=0$. This is in agreement with earlier results on Andreev transport in quantum dots,\cite{Trocha2014} and also in agreement with the corresponding results for layered hybrid structures.\cite{Jong}
The spin conductance (see Fig.~\ref{Fig:6}b) varies with $\varepsilon$ in a similar way as the single-particle contribution to charge current, and thus also in a similar way as the heat conductance, revealing two peaks associated with single electron tunneling. There is no contribution to the spin conductance due to Andreev tunneling, which is rather obvious as the net spin of a Cooper pair is zero, so Andreev current is not spin polarized.  The corresponding charge and spin thermopowers are comparable and display similar dependence on $\varepsilon /\Delta$, see Figs ~\ref{Fig:6}c,d. However, the corresponding figures of merit (see Figs.~\ref{Fig:6}e,f) are remarkably different, i.e. the spin figure of merit  is significantly smaller (roughly by one order of magnitude) than the conventional one. This is a consequence of the fact, that Andreev current is not spin polarized, so the spin polarization of the total compensating current is relatively small due to a significant contribution of the Andreev current.
\begin{figure}[]
\includegraphics[width=\columnwidth,angle=0]{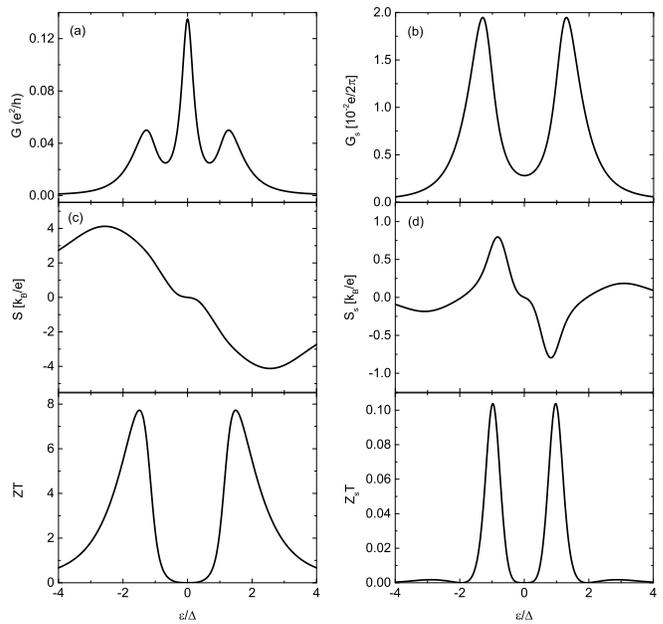}
\caption{\label{Fig:8}
Spin thermoelectric coefficients:
(a) charge conductance, (b) spin conductance, (c) thermopower, (d) spin thermopower,
(e) figure of merit $ZT$, and (f) spin figure of merit $Z_sT$,
calculated as a function of the dot's level energy for $k_BT/\Delta
=0.4$, $\Gamma/\Delta =0.1$, $U=0$, $p=0.9$ and
$r=2$.}
\end{figure}%

In Fig.~\ref{Fig:8} we show the same physical quantities  as in Fig.~\ref{Fig:6} but for a higher temperature, $k_BT/\Delta =0.4$, keeping the other parameters as in Fig.4. As in the case of nonmagnetic lead, the contribution from single electron tunneling is now clearly visible in Fig.~\ref{Fig:8}a as two well resolved side maxima. Variation of the spin conductance (Fig.5b)  with $\varepsilon$ is similar to that for $k_BT/\Delta =0.1$, except the conductance is significantly larger, compare Figs ~\ref{Fig:6}b and ~\ref{Fig:8}b. As the maximum value of the conventional thermopower (Fig.5c) is similar to that at lower temperature, the corresponding maxima in spin thermopower (Fig.5d) are remarkably decreased (roughly by a factor larger than 2). The corresponding conventional figure of merit (Fig.5e) is significantly larger, while the increase in the spin figure of merit (Fig.5f) is rather moderate in comparison to that in Fig.4f.

\section{Effects of finite $U$}

\begin{figure}[]
\includegraphics[width=\columnwidth,angle=0]{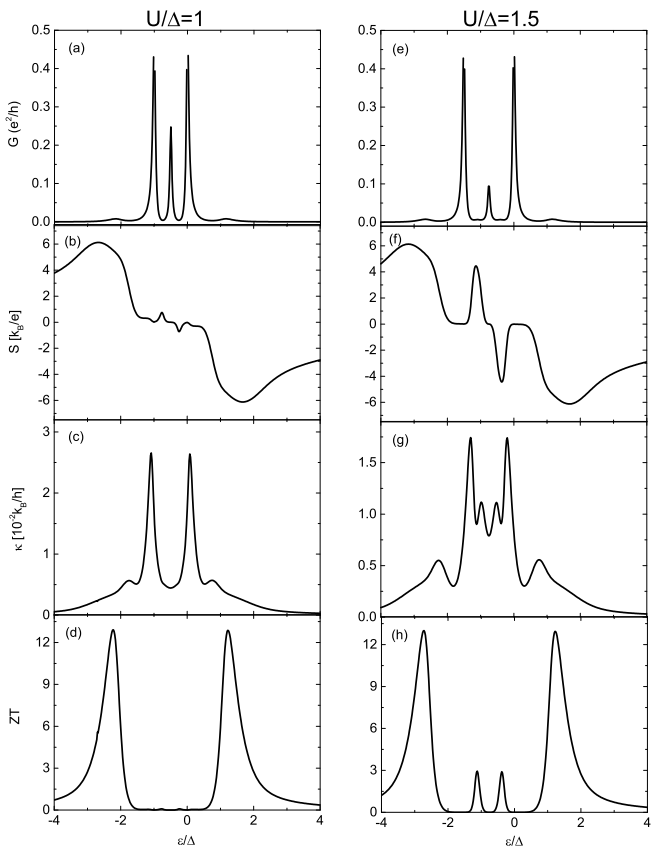}
\caption{\label{Fig:9}
Thermoelectric coefficients:
(a,e) charge conductance, (b,f) thermopower, (c,g) heat conductance,
and (d,g
) figure of merit $ZT$,
calculated as a function of the dot's level energy for $k_BT/\Delta
=0.2$, $\Gamma/\Delta =0.1$, $p=0$, $r=1$, and indicated values of $U/\Delta$.}
\end{figure}

Consider now the case of nonzero Coulomb  correlations in the dot, i.e. when $U$ is finite.
Let us begin with the case of nonmagnetic metallic lead, $p=0$.
The corresponding numerical results are shown in Fig.~\ref{Fig:9} for two different values of $U/\Delta$.
The electrical conductance reveals now three narrow peaks. These peaks correspond to Andreev tunneling. The side peaks occur at $\varepsilon =0$ and $\varepsilon =-U$, while the central peak appears in the middle between them (particle-hole symmetry point). Separation of the side peaks is equal to $U$, while their height is independent of the Coulomb parameter $U$. In turn, height of the central peak decreases with increasing $U$.
Apart from the three Andreev peaks, there is also a small contribution from tunneling of single electrons, see the weak side peaks   at $\varepsilon =\Delta$ and $\varepsilon =-\Delta-U$.

\begin{figure}[]
\includegraphics[width=\columnwidth,angle=0]{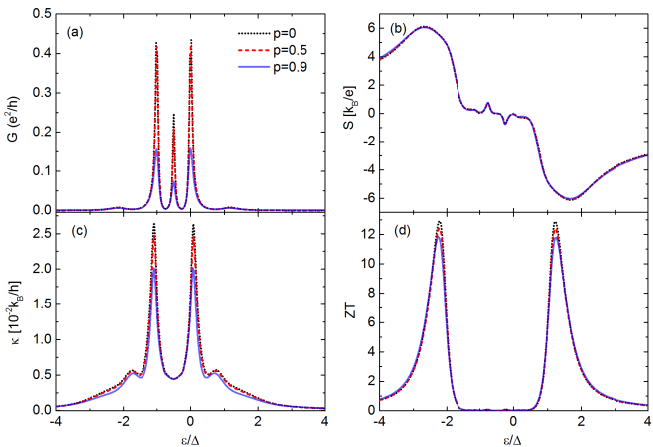}
\caption{\label{Fig:10}
Thermoelectric coefficients:
(a) charge conductance, (b) thermopower, (c) heat conductance, and
(d) figure of merit $ZT$,
calculated as a function of the dot's level energy for $k_BT/\Delta
=0.2$, $\Gamma/\Delta =0.1$, $r=1$, $U/\Delta=1$ and indicated values of $p$.}
\end{figure}%

The heat conductance [Fig.~\ref{Fig:9}(c,g)] as a function of $\varepsilon$ reveals not only the weak side peaks which may be attributed to
single electron transport, but also narrow and relatively high peaks which correspond to $\varepsilon$  where the Andreev peaks
in the charge conductance occur. Taking into account the fact that thermal conductance is determined on the condition of zero charge current,
one can note that these peaks follow from the interplay of tunneling of single electrons and Cooper pairs. Due to the more complex behavior of charge
conductivity with $\varepsilon$, the resulting thermopower is also more complex, as shown in Fig.~\ref{Fig:9}(b,f). The maximum values of the thermopower appear close to the side peaks in the electrical conductance and are roughly the same for both values of $U$ assumed in Fig.~\ref{Fig:9}. In turn, the figure of merit reaches maximum values close to the
points where $|S|$ has maxima. Interestingly, for larger $U$ some small peaks in $ZT$ appear between the high side peaks, which correspond to the  narrow maxima in $|S|$ close the particle-hole symmetry point.

Numerical  results for the case of magnetic lead in the absence of spin accumulation are shown in Fig.~\ref{Fig:10} for several values of the polarization factor $p$, while the other parameters are kept constant.
The charge conductance depends rather significantly on the polarization factor $p$, as follows from Fig.~\ref{Fig:10}a. This problem was already discussed above and is
due to suppression of Andreev tunneling with increasing $p$. Remarkably weaker dependence on $p$ takes place in the case of heat conductance (Fig.~\ref{Fig:10}b). This dependence appears mainly at the peaks which are associated with the peaks in Andreev tunneling, while the peaks associated with single electron tunneling are only weakly dependent on $p$. This behavior is a consequence of external voltage applied to suppress charge current when determining the heat conductance, as already described earlier. Similarly, the thermopower (Fig.~\ref{Fig:10}b) and figure of merit (Fig.~\ref{Fig:10}d) are weakly dependent on $p$, too.

\begin{figure}[]
\includegraphics[width=\columnwidth,angle=0]{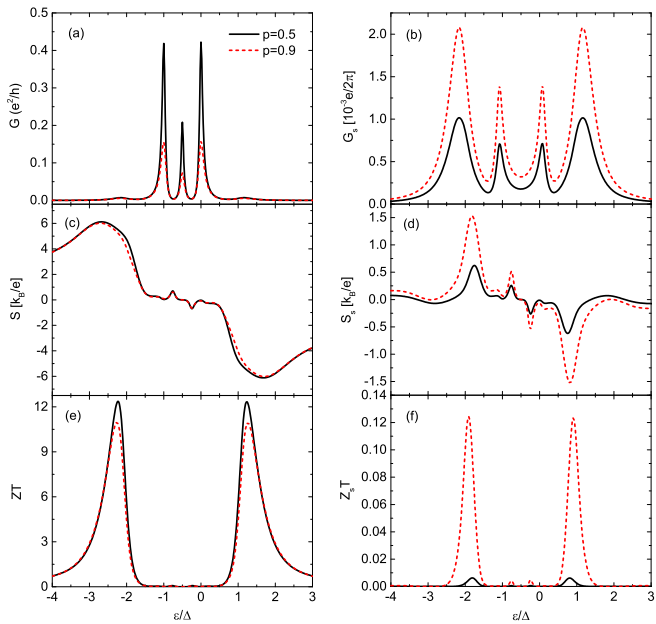}
\caption{\label{Fig:11}
Spin thermoelectric coefficients: (a) charge conductance, (b) spin conductance,
(c) thermopower, (d) spin thermopower,
(e) figure of merit $ZT$, and (f) spin figure of merit $Z_sT$,
calculated as a function of the dot's level energy for indicated values of spin
polarization $p$ and for $k_BT/\Delta
=0.2$, $\Gamma/\Delta =0.1$, $U/\Delta=1$, and
$r=2$.}
\end{figure}%

Let us now briefly discuss the thermoelectric parameters in the presence of finite
spin accumulation. In Fig.~\ref{Fig:11} we show these parameters  for
indicated values of the spin polarization factor $p$. First of all, the charge conductance reveals a strong response
on the change of spin polarization. As for the case of no spin accumulation (see Fig.~\ref{Fig:10}a) the conductance
decreases with increasing $p$ due to suppression of Andreev reflection processes discussed above.
The spin conductance acquires higher values for larger spin polarization factor, see Fig.~\ref{Fig:11}b, which is rather obvious as the spin conductance is a  measure of the difference in conductances of both spin channels. Opposite to the noninteracting case, the spin
conductance exhibits now peaks in the sub-gap region. Similarly, more complex behavior can be noticed
in the conventional thermopower (Fig.~\ref{Fig:11}c), and even more pronounced in the spin thermopower  (Fig.~\ref{Fig:11}d).
These features can be explained in the same way as earlier for the non-polarized case.
Moreover, the conventional thermopower only very weakly depends on the spin polarization factor,
similarly as in the case without spin accumulation. This does not hold for
the spin Seebeck coefficient which strongly varies with increasing $p$ factor. It grows with increasing $p$
and reaches a maximum value for half metallic lead, \ie for $p\rightarrow 1$ (not shown).
Obviously, both $G_s$ and $S_s$ vanish for $p=0$.
The corresponding spin analog of the figure of merit, $Z_sT$, shown in Fig.~\ref{Fig:11}f, grows
with increasing $p$, which results from the increase of $|S^s|$ and $G_s$ and a decrease
of the heat conductance $\kappa$ (not shown).
In turn, the figure of merit $ZT$, weakly decreases as $p$ grows.
This follows the corresponding dependence of $ZT$ calculated for the case without spin accumulation (see Fig.~\ref{Fig:10}
for comparison).


\section{Summary}

Using the nonequilibrium Green function method, we have analyzed transport and thermoelectric
properties of a quantum dot attached to one metallic (ferromagnetic in a general case) and one superconducting lead.
Coulomb correlations in the dot have been taken into account.
The main focus was on thermoelectric properties, like thermopower and figure of merit, including also spin thermopower and spin thermoelectric efficiency.
Our considerations were limited to the linear response regime, where we calculated electric, spin,  and thermal conductances as well as the thermpower, spin thermopower, figure of merit, and spin figure of merit. These characteristics, in general, depend remarkably on the relative temperature, i.e. temperature measured  with respect to the suerconducting gap.

It has been shown that single-particle transport to states above (below) the superconducting gap is responsible for pronounced thermoelectric properties, while Andreev tunneling itself creates neither heat current nor thermopower. However, there is a strong influence of the Andreev tunneling  (which contributing to the charge conductance) on the thermoelectric properties, which follow from the fact that both heat conductance and thermopower are defined at zero charge current. Thus, Andreev tunneling comes into play {\it via} the compensating current and leads to remarkable features in the heat conductance as well as in thermopower and figure of merit (including also spin thermopower and spin figure of merit).

Due to a strong increase in density of single-particle states near the superconducting gap edges, the thermopower as well as the figure of merit, both conventional and spin ones,  achieve quite remarkable values. Moreover,
it is also shown that Coulomb correlations in the dot have a significant impact on the transport and thermoelectric coefficients.

\begin{acknowledgements}
This work was supported by the National Science Center in Poland as the Project
No. DEC-2012/04/A/ST3/00372.
\end{acknowledgements}

\end{document}